\documentclass[
 reprint,
 amsmath,amssymb,
 aps, prb,
showpacs
]{revtex4}

\usepackage{amsmath}
\usepackage{amssymb}
\usepackage{textcomp}
\usepackage{float}
\usepackage{fmtcount}
\usepackage{hyperref}
\usepackage{cleveref}
\Crefname{figure}{FIG.}{FIGs.}
\usepackage{graphicx}
\usepackage{dcolumn}
\usepackage{bm}
\usepackage[bulgarian]{babel}
\usepackage[utf8]{inputenc}

\newcommand{\Fref}[1]{Фиг.~\ref{#1}}
\newcommand{\Eqref}[1]{Ур.~(\ref{#1})}

\begin{document}

\preprint{APS/123-QED}

\title{Planck's constant measurement by Landauer quantization for students laboratories\footnote{Published in Eur. J. Phys. 36 (2015) 055047 (13pp), \href{http://iopscience.iop.org/article/10.1088/0143-0807/36/5/055047/meta}{dx.doi.org/10.1088/0143-0807/36/5/055047}}}

\author{Desislav S. Damyanov, Iliana N. Pavlova, Simona I. Ilieva, Vassil N. Gourev, Vasil G. Yordanov, Todor M. Mishonov}
\email[E-mail: ]{\\desislav.s.damyanov@gmail.com,\\ ilinikpavlova@gmail.com,\\ simonailieva24@gmail.com, \\gourev@phys.uni-sofia.bg, \\vasil.yordanov@gmail.com, \\mishonov@gmail.com}
\affiliation{%
 Faculty of Physics, St.~Clement of Ohrid University at Sofia,\\ 5 J. Bourchier blvd., BG-1164 Sofia, Bulgaria
}%
\date{\today}

\begin{abstract}
A simple experimental setup for measuring the Planck's constant, using Landauer quantization of the conductance between touching gold wires, is described. It consists of two gold wires with thickness of $500 \,\mathrm{\mu m}$ and $1.5\,\mathrm{cm}$ length, and an operational amplifier. The setup costs less than \$$30$ and can be realized in every teaching laboratory in two weeks. The usage of oscilloscope is required.
\end{abstract}

\pacs{01.50.Pa,
73.63.Nm,	
73.63.Rt	
}
\keywords{Planck's constant, Planck's constant measurement, conductance quantum}
\maketitle


\section{\label{sec:intro}Introduction}

The purpose of the present work is to describe an experimental setup for observation the Landauer's conductance quantization and determination of Planck's constant \cite{Planck}, which can be easily realized in every teaching laboratory.

The quantization of the conductance was predicted by Rolf Landauer in 1957 \cite{IBM}. See, for example, Landauer's \cite{Landauer1989} and B\"uttiker's \cite{Buttiker} surveys.

In the last years the observation of the conductance quantum and its usage for determining the Planck's constant has become an ordinary experimental work for students \cite{Tolley,Foley,conference,Soukiassian}.

In order to receive information about the history of measuring the Planck's constant check, for instance, the Steiner's survey \cite{Steiner}.

The conductance quantization has an elementary explanation, suitable for high-school students which can be considered as an illustration of Bohr's model of hydrogen atom \cite{Stoev}.

The experimental setup is available in the Atomic physics laboratory. As described in other works \cite{Tolley,Foley}, it is up to the students' imagination to find a way to connect and disconnect the wires. However, the experimental setup is robust and repeatable in the time frame of three hours (the duration of the laboratory class). If break junctions are used \cite{Soukiassian}, the measurement becomes more reliable but the flair of the experimental fineness is lost.

The most reproducible experiment on conductance quantization is conducted in technological structures with no atomic movement. The gain voltage is the only regulator of the quantum channels \cite{Wees,Kouwevhoven,Kouwenhoven1992}.


\section{\label{sec:theor}Theoretical model}


\subsection{\label{sec:theo_1D}Conductance of one dimensional metal excluding scattering}
\begin{figure}
\includegraphics[width=0.47\textwidth]{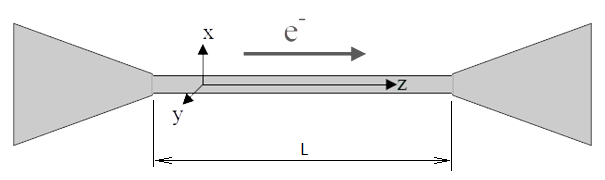}
\caption{Landauer picture for one dimensional transport\cite{IBM,Landauer1989}. One dimensional (1D) electron waveguide between two massive metallic electrodes. Different transversal modes correspond to different quantum channels of conductivity\cite{Kouwenhoven1992}. The mean free path is bigger than waveguide length $L$ and the electrons fly ballistically trough the contact area between the electrodes. In the experimental realization of the present paper the electrodes are two gold wires, while the waveguide is just the point contact between them cf. \cite{Tolley}.}
\label{Landauer_1D_conductor}
\end{figure}

According to the Landauer consideration\cite{IBM,Landauer1989} a conductor can be assumed as a one dimensional system of free electrons schematically presented in Fig.~\ref{Landauer_1D_conductor} if length $L$ is much smaller than mean free path.

The distribution of the electrons over the energy states in a system with a large number of identical particles, is described by Fermi-Dirac's statistics\cite{Fermi}. Fermi-Dirac's distribution gives the average number of fermions with momentum $p$ in a single-particle state\cite{Reif}
\begin{equation}\label{}
n_p=\frac{1}{exp[(\epsilon _p - \mu)/k_BT^{\prime}]+1},
\end{equation}
where $k_\mathrm{B}$ is the Boltzmann's constant, $T^{\prime}$ is the absolute temperature, $\epsilon _p$ is the energy of the single-particle state, supposedly equal to $\epsilon _p=p^2/2m$, $\mu$ is the chemical potential and $p$ is the electron momentum. At zero temperature the chemical potential is a sum of the Fermi energy and the potential energy per electron\cite{Blakemore,Kittel}.

There are two possible values for $n_p$

\begin{equation}\label{}
n_p=\begin{Bmatrix}
1\textrm{ in case of }\epsilon _p < \mu\\ 
0\textrm{ in case of }\epsilon _p > \mu\\ 
\end{Bmatrix},
\end{equation}
which derive from Pauli's principle. 

Now we can present the current as a flux of electrons using summation
\begin{equation}\label{}
I=q_e\sum_{\alpha,p;\;(v_p>0)}v_p\frac{\bar{n_p}}{L},\qquad v_p=\partial _p\epsilon _p=\frac{p}{m},
\end{equation}
where $q_e,\alpha$ and $v$ respectively are: the charge, spin and velocity of the electron. We suppose that $q_e$ is known. 
The magnitude $\bar{n_p}/L$ is averaged space density of electrons, having momentum $p$ and $q_e\bar{n_p}/L$ is the electrical density and represents the average number of particles per unit length.
Implicitly in the Landauer consideration is supposed ballistically propagation of free particles in periodic boundary conditions with geometrical period $L.$

We can transform the summation into integration using phase integral \cite{Laurendeau}
\begin{equation}\label{}
\sum_{p}=\frac{L^D}{(2\pi\hbar)^D}\int_{0}^{p_F}d^Dp,\qquad v=\frac{p}{m}\geq 0,
\end{equation}
where $D$ is the space dimension, $D=1$ is our case, $\hbar$ is the reduced Planck's constant, $L^D$ is the considered volume (in this case it is the length of the conductor) and $p_F$ is the Fermi momentum.
Formally summation can be substituted by integration only in $L\rightarrow\infty$ limit.

Therefore, the current can be written as
\begin{equation}\label{eq:current}
I=2q_eL\int_{0}^{p_F}\frac{p}{m}\frac{dp}{2\pi \hbar}\frac{1}{L},
\end{equation}
where the first multiplier 2 takes into account spin summation. The velocity is presented as $p/m$ and $h=2\pi\hbar$.

After integration, the equation becomes
\begin{equation}\label{}
I=\frac{2q_e}{2\pi \hbar}\left (\frac{p^{2}}{2m}\right)\bigg\vert_{0}^{p_F}=\frac{2q_e}{2\pi \hbar}E_F=\frac{2q_e}{2\pi \hbar}q_eU=\frac{2q_e^{2}U}{h}=\sigma_0 U,
\end{equation}
where $\sigma_0$ is quantum of conductance \cite{Soukiassian}
\begin{equation}
\label{eq:sigma}
\sigma_0 = \frac{q_e^2}{\pi \hbar}
=\frac{2}{R_\mathrm{H}}
=77.5\,\mathrm{\mu S}=\frac{1}{12.9 \,\mathrm{k\Omega}},
\end{equation}
where $R_\mathrm{H}=2\pi\hbar/q_e^2=25812.8074434(84)\,\Omega$ \cite{Mohr}
is the quantized Hall resistance  (von Klitzing constant), related to the metrological definition of the resistance unit $\Omega.$
The Planck's constant is defined assuming that the elementary charge is given.
It is remarkable that in this result \eqref{eq:sigma} effective mass of quasiparticles $m$ is canceled and tis result is applicable for all 1D conductors. The length $L$ is also irrelevant and conductivity quantization can be seen not only in 1D electron waveguides, but even for point contacts for which $L=0$, see~\cite{Landauer1989,Kouwevhoven}. For applicability of Landauer quantization for touching wires is necessary size of the contact (of order of few \AA) to be much smaller than the mean free path in metals which is in order of 100~\AA. The electrons have to fly trough the contacting area as free particles. That is why thin oxide layer in copper can smear the conductivity quantization and it is better to use gold wires. 

The conductance quantum is also related to the Bohr's velocity
\begin{eqnarray}
&&\frac{1}{4\pi \epsilon _0R_\mathrm{H}}
=\frac12\,\frac{\sigma _0}{4\pi \epsilon _0} = \frac{v_\mathrm{_{Bohr}}}{2\pi}, \\
&&v_\mathrm{_{Bohr}} = \frac{e^2}{\hbar }
=\alpha_\mathrm{_{Zommerfeld}}c,
\quad e^2 = \frac{q_{e}^2}{4\pi\epsilon_0}.\nonumber
\end{eqnarray}
All formulae are written in SI. 
In CGS system $4\pi\epsilon_0=1$, and the conductance has its natural  dimensionality velocity \cite{Parsell}. In Heaviside-Lorentz system $\epsilon_0=1=c$.

There is a simpler method to calculate the integral for the current in Eq.~\eqref{eq:current} using only summation of arithmetic progression~\cite{Stoev}, which is suitable for high-school students.


\subsection{\label{sec:landauer}Landauer formalism}
A more realistic case in observing the current flowing through a conductor is when scattering is taken into account. 

Consider the conductor as a two dimensional system. The electrons are confined by an infinite 2D (two dimensional) potential in x and y directions. In this case we should use quantum mechanics, in particular the Schr\"odinger equation for a particle in a box. Its solution gives discrete number of eigenstates, which are also called modes (or channels if conduction is considered). The total energy of the conductor can be obtained as a sum of the lateral mode energy and the energy of the one dimensional solution in the z-direction.

Using the Landauer formalism for a current flow through small constrictions, the following can be derived

\begin{equation}
\label{}
\sigma =\sigma _0\sum_{i=1}^{N}T_i,
\end{equation}
where $\sigma _0$ is the quantum conductance, $N$ is the number of conduction channels and $T_i$ is transmission coefficient for the $i^{th}$ channel in the wire \cite{Landauer1989}\cite{Soukiassian}. The factors $T_i$ are significant since the conductor is not ideal. They represent the probability that an electron will traverse the constriction, traveling through the $i^{th}$ channel. $T_i$ differs from $1$ when backscattering becomes important in the transport process. If the length of break junctions is smaller than the mean free path of an electron in a metal ($\sim 100\,\textrm{\AA}$), it can be accepted that $T_i=1$ for all channels. The Landauer formula becomes
\begin{equation}\label{}
\sigma =\sigma _0N\textrm{  or } \frac{\sigma }{\sigma _0}=N,
\end{equation}
i.e. the conductance divided by quantum conductance is always equal to an integer \cite{Soukiassian}.

In the next section we describe the experimental realization.

\section{\label{sec:experiment}Experimental part}
\subsection{\label{sec:el_circuit}Electric circuit}

\begin{figure}[h]
\includegraphics[width=0.6\textwidth]{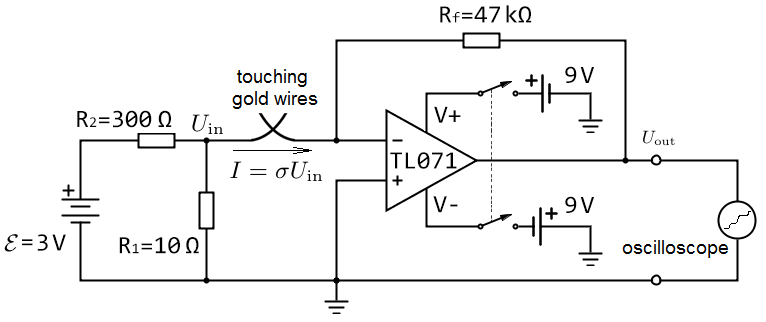}
\caption{
Electrical scheme for measurement of Plank's constant. The current to voltage converter formed by the operational amplifier TL071 and the feedback resistor $R_\mathrm{f}$ has output voltage proportional to the input current $U_\mathrm{out}=-R_\mathrm{f} I$. The input current that flows through the gold wires is given by the conductivity $I=\sigma U_\mathrm{in}$, where $U_\mathrm{in}$ is the voltage drop on $R_1$ from the voltage divider form by $R_1$ and $R_2$. The output voltage is proportional to the conductivity of the gold wires $U_\mathrm{out}=-\sigma R_\mathrm{f} U_\mathrm{in}$ and it is recorded with digital oscilloscope. 
An alternative analysis of the circuit is that input voltage $U_\mathrm{in}=\mathcal{E}R_1/(R_1+R_2)$ is amplified by an inverting amplifier with
the feedback resistor $R_\mathrm{f}$ and input resistor $1/\sigma$. The amplification coefficient is $G(t)=-\sigma R_\mathrm f$ and the oscilloscope shows the time dependent output voltage $U_\mathrm{out}(t)=G U_\mathrm{in} \propto \sigma(t).$ In such a way we can see the quantized time dependent conductivity between touching gold wires. 
}
\label{fig:schema}
\end{figure}

The experiment is conducted using the electric circuit presented on FIG.~\ref{fig:schema}, where crossed lines represent two gold wires. This circuit is similar to the one used in \cite{Foley}.

For observing the time-dependence of the conductance and its quantization, the current through the conducting wires is measured using a current to voltage converter, as it is shown on FIG.~\ref{fig:schema}.

When the gold wires form one quantum of conductance, they can be presented as a resistor $r_1=12.9 \,\mathrm{k\Omega}$. In this case, FIG.~\ref{fig:schema} shows the scheme of an inverting amplifier, which measures the input voltage $U_\mathrm{in}$ from the voltage divider formed by $R_1$ and $R_2$.
For the values of $R_1=10\, \mathrm{\Omega}$ and $R_2=300\, \mathrm{\Omega}$ the voltage drop on $ R_1$ of the divider is approximately $100\, \mathrm{mV}$. The value of $R_1$ is chosen in a such way that the voltage divider is not loaded by the input resistance of the inverting amplifier for the first tens of the quantum resistances $R_1 \ll r_N, N=1,\dots, 10$. The value of $R_2$ is chosen so that the output voltage from the divider $U_\mathrm{in}$ is less than the maximum output voltage of the inverting amplifier, divided by its gain $G_n$ for the first tens of the quantum conductance $U_\mathrm{in} = 3\, \mathrm{V} \times R_1/(R_1+R_2) < 9\, \mathrm{V} / G_N , N=1,\dots, 10$. 

The voltage gain of an inverting amplifier is given as
\begin{equation}\label{}
G_N=\frac{U_\mathrm{out}}{U_\mathrm{in}}=-\frac{R_\mathrm{f}}{r_N}
\end{equation}
The used op-amp is TL071. It is very cheap, widespread and available in Dual In-line Package (DIP). It has bandwidth $3 \,\mathrm{MHz}$ corresponding to gain $G=1$. In the case of one quantum of conductance the gain is $G_1 = \frac{R_\mathrm{f}}{r_1}=3.7$. Since the product of the bandwidth of an op-amp with its gain is constant, the bandwidth of the used inverting amplifier is $B=0.8 \,\mathrm{MHz}$. This is much smaller than the bandwidth of the used oscilloscopes \cite{DS5042M,DS1052E}, which makes them appropriate for measurement of the output voltage $U_\mathrm{out}$.
The bandwidth of the inverting amplifier is also high enough, so that we can measure the first few quantum steps with length greater than $1/B=1.25\, \mathrm{\mu s}$. This is the reason to choose relatively small value of $R_\mathrm{f}$ is of order of several tens of $\mathrm{k\Omega}$, so that the gain is also kept small and respectively the bandwidth is high.
The aim of the experiment is to determine the Planck's constant using the values of quantized conductance between two touching gold wires. 
This conductance, and corresponding resistance are related to the Planck's constant 
\begin{equation}\label{r_N}
r_N=\frac{1}{\sigma }=\frac{1}{N\sigma _0}=\frac{1}{2q_e^2} \frac{h}{N}.
\end{equation}

The resistance $r_N=-\frac{U_\mathrm{in}}{U_\mathrm{out}}R_\mathrm{f}$ is obtained by measuring the value of $U_\mathrm{out}$, corresponding to the $N$\textsuperscript{th} quantum level, and taking into account the values of the input voltage $U_\mathrm{in}$ and the feedback resistance $R_\mathrm{f}$. Therefore, the Planck's constant is
\begin{equation}\label{eq:hconst}
h = 2\,q_e^2\,r_N N = -2\,q_e^2\,R_fN \,\frac{U_\mathrm{in}}{U_\mathrm{out}}.
\end{equation}
It is expressed by physical constants and experimentally determinable parameters and variables of the setup.
\begin{figure}
\includegraphics[width=0.6\textwidth]{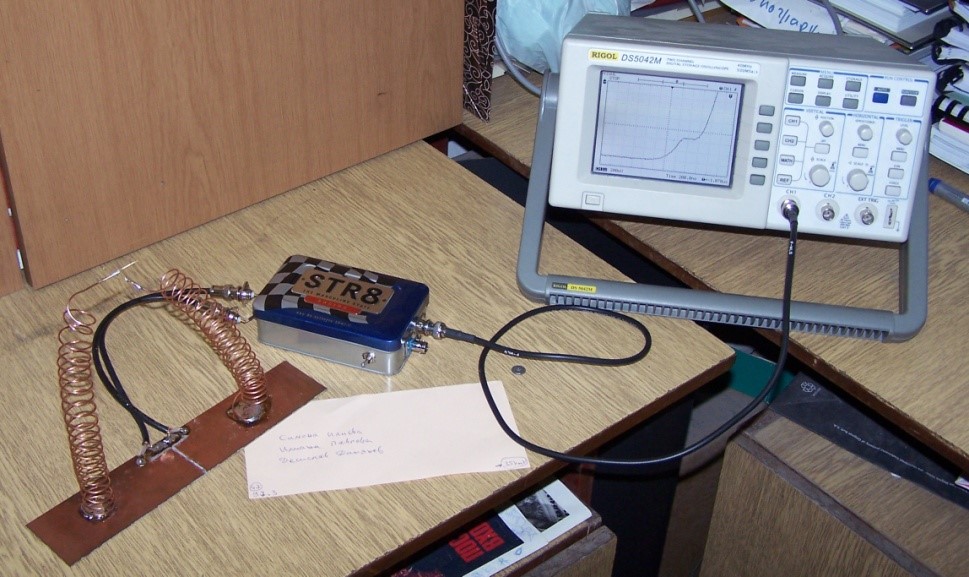}
\caption{
View of whole
 experimental setup.
One can see the oscilloscope showing a typical quantized step of the conductivity.
The electronic part is in a metallic box shown with open cover in great details in FIG.~\ref{fig:electric}.
The mechanical part ensuring slow touching of gold wires is given with great magnification in 
FIG.~\ref{fig:mechanic}.
All parts of the setup are connected with coaxial cables terminated by BNC connectors.
}
\label{fig:setup}
\end{figure}

\subsection{\label{sec:level2}Construction of the setup}
\begin{figure}
\includegraphics[width=0.47\textwidth]{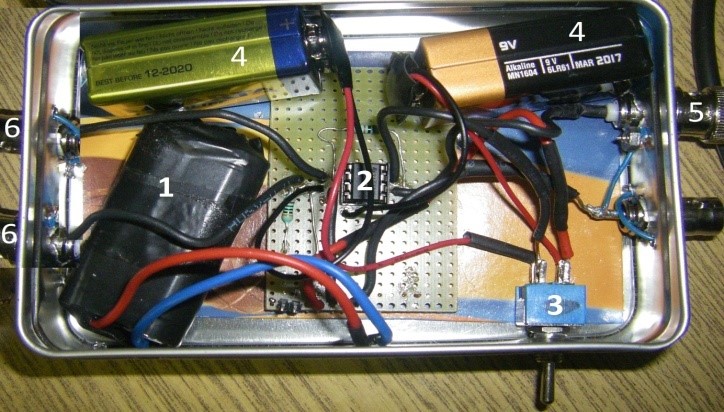}
\caption{Electrical circuit in the metallic box of Fig.~\ref{fig:setup}. 
One can see: 
1) wrapped by black isolating band the two AA batteries of 1.5 V each giving electromotive force $\mathcal{E}$,
2) operational amplifier (op-amp) TL071 \cite{TL071}, 
3) double switch to the voltage supply of the op-amp,
4) two 9~V batteries of the voltage supply of the op-amp,
5) BNC connector to the oscilloscope
6) BNC connectors for coaxial cables connected to the gold wires given in great magnification in FIG.~\ref{fig:mechanic}.
One can see also: the isolator plate on the bottom of the metallic box,
standard circuit board with holes,
on the left of the circuit board one of the resistors of the voltage divider,
feedback resistor $R_\mathrm{f}$ above the op-amp
and some unessential details. 
}
\label{fig:electric}
\end{figure}

\begin{figure}
\includegraphics[width=0.47\textwidth]{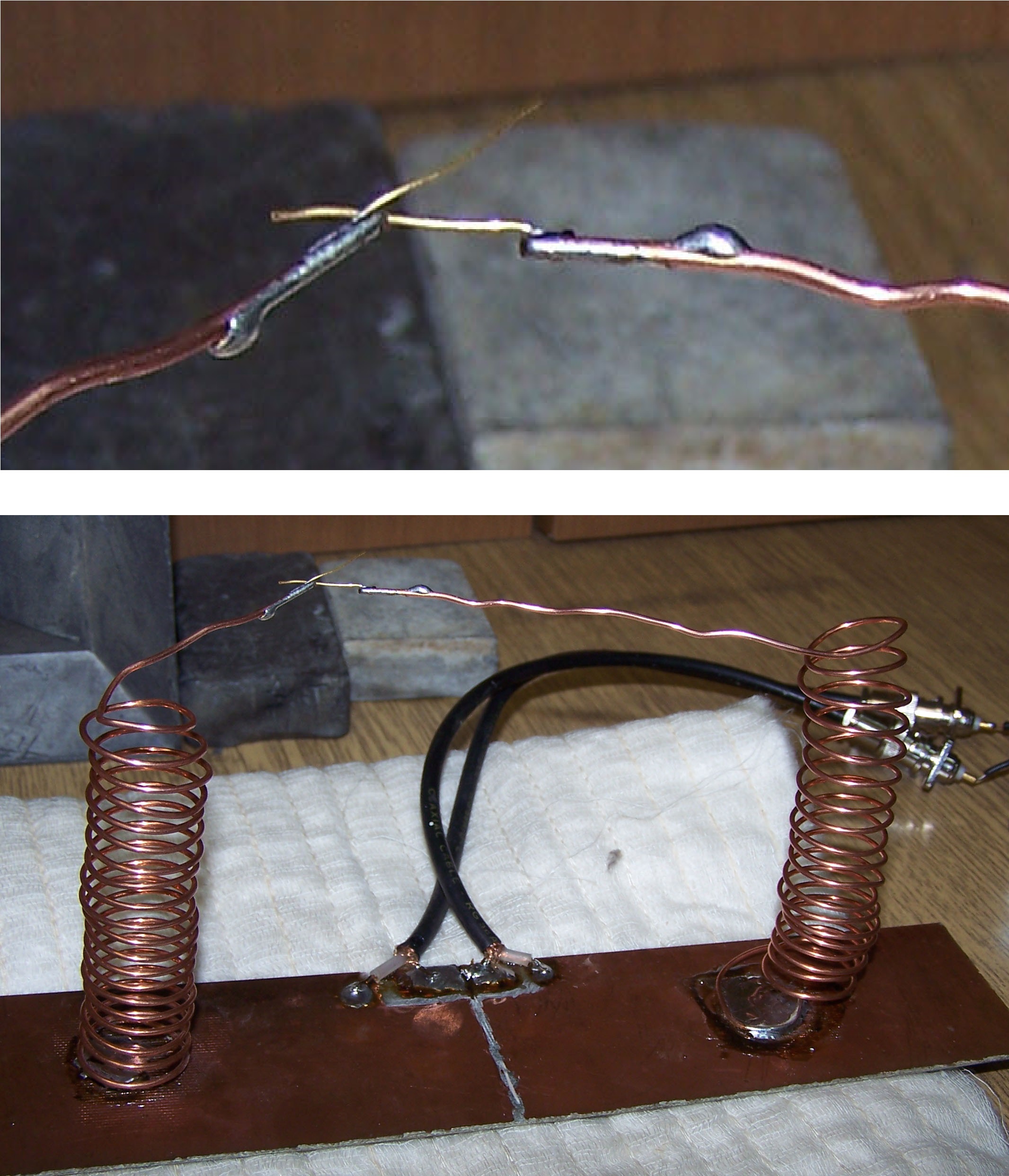}
\caption{The mechanical part of the setup. 
Short gold wires 
($15 \,\mathrm{mm}$ long and $0.5 \,\mathrm{mm}$ thick) 
are soldered at the top end of $1.5 \,\mathrm{mm}$ thick copper springs. 
The gold wires (zoomed in the upper part of the figure) are in mechanical contact as it is schematically presented at Fig.~\ref{fig:schema}.  
The bottom end of the springs are soldered on separated copper plates. 
To the copper plates are soldered wires of two coaxial cables, which reduce the external electric noise.
The shields of coaxial cables are connected to the common point (ground) of the scheme in Fig.~\ref{fig:schema}.
The purpose of the spring shaped copper wires is to decrease the frequencies of the eigen modes.
This is an mechanical filter for high frequency mechanical noise and simultaneously increase the time for observation of quantized steps of the conductivity.
When these steps are long enough we can use low frequency operational amplifiers and oscilloscopes. 
}
\label{fig:mechanic}
\end{figure}

The setup on FIG.~\ref{fig:setup} consists of a mechanical and an electrical part, which are connected with coaxial cables. The electrical and the mechanical parts are shown on FIG.~\ref{fig:electric} and FIG.~\ref{fig:mechanic} respectively.

The mechanical part is constructed in such a way in order to provide slow movement of the gold wires when they form or destroy an electric contact.
The gold wires are soldered at the end of big springs (see FIG.~\ref{fig:mechanic}). The wires are made of commercial 24-karat gold for jewelry. They are $15 \,\mathrm{mm}$ long and have a thickness of $0.5 \,\mathrm{mm}$. When the springs oscillate, the gold wires touch and detach. 
Each spring is $1.5 \,\mathrm{mm}$ copper wire with $9 \,\mathrm{cm}$ length. It has $20$ turns with a diameter of $2 \,\mathrm{cm}$. 
The springs are soldered on a cooper board with split layers in the middle to form two individual conducting surfaces. 

The electric circuit described in Sec.~\ref{sec:el_circuit} is implemented in the blue metal box shown on FIG.~\ref{fig:electric}. The DC voltage source is a $3 \,\mathrm{V}$ battery cell consisting of two batteries of $1.5 \,\mathrm{V}$ each. 
The operational amplifier TL071 is supplied by two batteries of $9 \,\mathrm{V}$. The switch breaks the power supply of the op-amp. 

The electrical part is connected with the mechanical part and the oscilloscope with $30 \,\mathrm{cm}$ long coaxial cables. The BNC connectors on the box match the characteristic impedance of the cables, which is $50 \,\mathrm{\Omega}$. 

The blue metal box of STR8 serves as a Faraday cage to suppress electromagnetic interference from the environment.

\subsection{\label{sec:level3}Measurement} 

In the beginning, the switch must be turned on to power the op-amp. Then, we set the oscilloscope's time scale, voltage scale and trigger level (for further details see Sec.~\ref{sec:guidelines}).

After a little push, the spring-shaped conductors start to vibrate and the gold wires come in and out of contact. When the wires connect, electrons pass from one of the wires to the other through $N$ opened quantum channels (see Sec.~\ref{sec:landauer}). 
Simultaneous to the detachment of the wires, the number of modes decreases and so does the output voltage. The resulting signal is a sequence of steps, due to the conductance quantization of the wires. The height of each step corresponds to an integer quantum of conductance. $U_\mathrm{out}$, corresponding to each step, is measured using the oscilloscope. The input voltage of the inverting amplifier $U_\mathrm{in}$ is measured directly with a voltmeter as a voltage drop on the resistor $R_1$. The resistance $R_\mathrm{f}$ is measured with an ohmmeter. Then, Plank's constant is calculated using Eq.~\eqref{eq:hconst} and the measured values of  $U_\mathrm{out}$, for a chosen $N$\textsuperscript{th} step, $U_\mathrm{in}$ and $R_\mathrm{f}$.

When performing the experiment, two approaches are possible.
The first approach is to separate the gold wires. The second approach is to put them in contact.
 
In the first approach, if the wires are connected in the beginning of the measurement, the output voltage is around $-9 \,\mathrm{V}$, because $r_N=0$ and the amplification of the inverting amplifier is close to its open-loop gain and $U_\mathrm{in}$ is amplified up to the supplying voltage.  In this case the oscilloscope's trigger level is set at around $700 \,\mathrm{mV}$ below it. The conductance quantization is observed during the separation of the wires.

In the second approach, if the wires are initially detached, the output voltage is around $0\, \mathrm{V}$, because the gain of the inverting amplifier is zero and also it is disconnected from the input source. In this case the trigger level is set at approximately $700 \,\mathrm{mV}$ above the zero level, which corresponds to the output voltage $U_\mathrm{out}$ when there are several quantum of conductance. The conductance quantization occurs when the wires touch together.

Both methods are appropriate for measurements and seem to be equally effective.

\section{\label{sec:result}Results}

The input voltage is measured to be $U_\mathrm{in}=97.3\,\mathrm{mV}$ and it does not change during the experiment. The measured value of $R_\mathrm{f}=(47.0\pm0.1)\,\mathrm{k}\Omega$. 
Accuracy of of this value is much higher than the accuracy of measurements by oscilloscope, that is why the uncertainty of $R_\mathrm{f}$ is irrelevant. 

\begin{figure}
\includegraphics[width=0.6\textwidth]{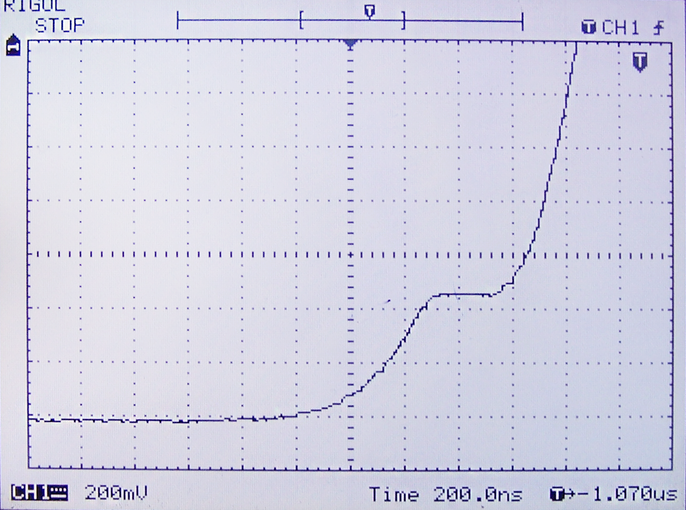}
\caption{Voltage proportional to the conductivity as function of time. One can easily see a step at which the conductivity is equal to the conductivity quantum $\sigma_0.$ Regime of disconnection of the gold wires. A $40 \,\mathrm{MHz}$ bandwidth oscilloscope is used. The X-axis is set to $200\,\mathrm{ns}$ and the Y-axis is at $200 \,\mathrm{mV}$. The step with height of $440\,\mathrm{mV}$ and length of $200 \,\mathrm{ns}$ corresponds to a single quantum conductance unit. The width of the step with quantized conductance is around 200~ns.}
\label{fig:conductance_step_a}
\end{figure}

\begin{figure}
\includegraphics[width=0.6\textwidth]{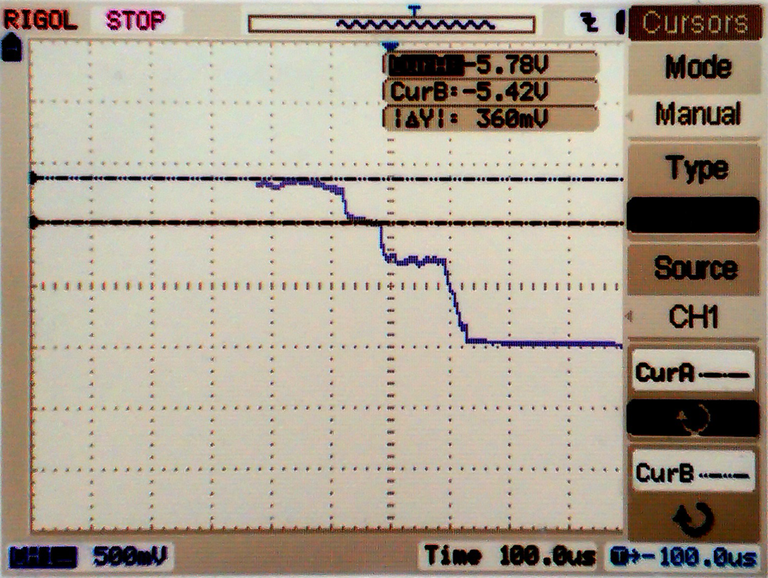}
\caption{Three steps of quantised conductivity as function of time can bee seen in regime of connection of the gold wires. A $50 \,\mathrm{MHz}$ bandwidth oscilloscope is used. The X-axis is set to $100\,\mathrm{\mu s}$ and the Y-axis is at $500 \,\mathrm{mV}$. The step between the markers with height of $360\,\mathrm{mV}$ and length of $60 \,\mathrm{\mu s}$ corresponds to a single quantum conductance unit. The horizontal cursors describes the chosen level of the conductivity. The variation of the curve from the horizontal step of quantized conductivity is more informative than an error bar around an experimental point. Note that the width of the steps $50-100\,\mu$s is significantly bigger than the width of the step depicted in \Cref{fig:conductance_step_a}.}
\label{fig:conductance_step_b}
\end{figure}

~\Cref{fig:conductance_step_a,fig:conductance_step_b} present photographs of the oscilloscope's display. They illustrate the op-amp output voltage as a function of time, using two different oscilloscopes.~\Cref{fig:conductance_step_a} show the experimental curves, when measurement is performed in regime of separation of wires, using the digital oscilloscope Rigol DS5042M. 
The curves on~\Cref{fig:conductance_step_b} are displayed in regime of connection, using digital oscilloscope Rigol DS1052E. The number of the open quantized channels is matter of interpretation. We approximate to the nearest integer number $N.$
Results from the measurements are presented in TABLE~\ref{fig:table}.
\begin{table}[!ht]
\centering
\begin{center}
\begin{tabular}{@{\hspace{10pt}}c@{\hspace{5pt}}c@{\hspace{5pt}}c@{\hspace{5pt}}}
\hline
\hline
  output  & number  & Planck's  \\
  voltage & of channels & constant\\
   $U_\mathrm{out}$ [mV]   & N & {$h\,[10^{-34}$J\,s]}   \\
\hline
 720 & 2 & 6.522 \\
 760 & 2 & 6.179 \\
 1800 & 5 & 6.522 \\
 360 & 1 & 6.522 \\
 440 & 1 & 5.336 \\
 360 & 1 & 6.522 \\
 320& 1 & 7.337 \\
 360& 1 & 6.522 \\
 320& 1 & 7.337 \\
 720& 2 & 6.522 \\
 700& 2 & 6.708 \\
 360 & 1 & 6.522 \\
 360 & 1 & 6.522 \\
 360 & 1 & 6.522 \\
 360 & 1 & 6.522 \\
\hline
\hline
\end{tabular}
\caption{Results from the measurements obtained from different experiments. The first column is the voltage level of the conductance step with N channels. See for example~\Cref{fig:conductance_step_a,fig:conductance_step_b}. The number of channels is written in the second column. In the last column is given the calculated value of the Planck's constant.}\vspace*{12pt}
\label{fig:table}
\end{center}
\end{table}

Planck's constant evaluated as a mean value and standard deviation of our measurements represented in TABLE~\ref{fig:table} is
\begin{equation}\label{}h=(6.54\pm  0.45)\times 10^{-34} \, \mathrm{J\, s}\,.\end{equation}
The deviation from the CODATA 2010 recommended value ($h = 6.626\,07123(133)\times 10^{-34}\,\mathrm{J\,s}$) \cite{Mohr} is 1.3\%.

\section{\label{sec:guidelines}Guidelines for students and laboratory teaching assistants}

In order to estimate Planck's constant using the method and experimental setup presented in this work, students should follow the instructions below.

Turn on the switch on the metal box. The switch breaks the voltage supply of the op-amp. Then, measure the voltage immediately before the gold wires, using a voltmeter. 
The measured value is the input voltage $U_\mathrm{in}$ and is approximately $100 \,\mathrm{mV}$.

After that, connect the oscilloscope to the electrical part of the setup, using the BNC's.
We use oscilloscopes Rigol DS5042M \cite{DS5042M} and Rigol DS1052E \cite{DS1052E}, but any digital oscilloscope with sampling rate bigger 10 mega-samples per seconds can be used. This frequency is determined by the time interval of the quantized conductivity. 
For our setup those times are of order $\mu$s, and in case of good mechanics with very low eigen frequencies time intervals can be even bigger up to 1~ms.

Set the time scale of the device. Take into account that the stability of the mechanical part of the setup influences the voltage dependence on time. 
In case of slow connection and disconnection of the gold wires the $X$ scale should be set to 
$100 \,\mathrm{\mu s/div}$. Otherwise, the $X$ scale should be $0.2-50 \,\mathrm{\mu s/div}$.

Set the voltage scale. The gain of the inverting amplifier is $G_N=-\frac{1}{r_N}\,R_\mathrm{f}=3.7\,N$ for $N$ conductance quanta. As the resulting voltage is  $U_\mathrm{out}=-G_N\,U_\mathrm{in}$, the voltage scale should be set to $200-500 \,\mathrm{mV/div}$.

Furthermore, a quick check on the reliability of the setup is recommended. When the gold wires are steadily connected, a horizontal line at around $-\,9 \mathrm{V}$ appears on the oscilloscope's screen. The resulting voltage is approximately equal to the supply voltage of the inverting op-amp. When the wires are separated from one another, the resulting voltage is $0\,\mathrm{V}$ and a horizontal line at $0$ should be on display. 

At the start of each measurement the wires can be connected or divided. As the resulting signal is supposed to be a sequence of steps, a trigger level must be set. Appropriate trigger mode is "Edge". An edge trigger occurs when the trigger input passes through a specified voltage level at the specified slope direction. The trigger sweep should be "Single". This way, whenever a trigger event occurs, the oscilloscope acquires one waveform and stops.

If the wires are initially connected, the triggering is on rising slope. In case they are divided, the triggering is on failing slope.  The trigger level should be set at around $U_\mathrm{trig}=U_\mathrm{out} (N=2)\approx 700 \,\mathrm{mV}$ below or above the initial line for the two approches respectively. Teaching assistants can find more information on triggering in the oscilloscope's user manual\cite{DS5042M,DS1052E}.

When performing the experiment, the springs can be shaped in different ways and various actions can be taken to improve the resulting signal. It is possible to resize one or both of the springs, restrict their movement, for example make one of them still, tilt the cooper plate, put a soft fabric under the plate, make bigger springs with lower vibration frequency and many more. Students are given a chance to come up with and test their own ideas and moderate the springs the way they prefer. For example, students from the same course made alternative realization, using long lever with attached micro-metric screw on its long arm. When they turn the screw, the short arms moves a gold wire that touches or detaches another gold wire.

When the desired motion manner is achieved, a little push on the side of the cooper plate or on the surface of the experimental table would be sufficient to provide good signal on the oscilloscope's screen. In most cases, many attempts are made before the goal is reached. It is also very likely that only one step is clearly observable on the oscilloscope. This means that several graphs should be obtained in order to estimate Planck's constant accurately. Once the graphs are obtained, the height of each step should be calculated. This is the output voltage $U_\mathrm{out}$. 

After all the measurements are done, the switch must be turned off in order to disconnect the op-amp from the power supply.

Finally, the Planck's constant is calculated using equation \ref{eq:hconst}.
The CODATA 2010 value of the elementary charge is $1.602176565(35)\times 10^{-19} \,\mathrm{C}$ \cite{Mohr}. Also, the electron charge can be measured in teaching labs. For this purpose, a setup, made by students, is available at the Atomic physics teaching laboratory.
The estimated value of the Planck's constant should be compared to the CODATA 2010 recommended value.

\section{\label{sec:conclusion}Conclusion}
Being ubiquitous in our everyday life, it is hard to believe how handy the contact between metal conductors can be for measuring a fundamental physics constant. The Landauer quantization happens to be a surprisingly good basis for a true experiment, leading to miraculously good results for such a simple experimental setup. The opportunity to observe quantum conductance units as steps in the time dependence of voltage, produced by the fine touch between two wires, is remarkable. It took years to measure Planck's constant with high accuracy, but within 2\% accuracy for a teaching laboratory having an oscilloscope, \$30, couple of weeks and a small group of motivated students are all the resources you need. Undeniably, this price is accessible for every teaching laboratory with the desire to make noticeable achievements by its simplicity and effectiveness. Since different approaches are possible, a dose of enthusiasm and preciseness may become the fuel for a valuable accomplishment, making it worthy for the teaching assistant to participate in. 

In order to pass the exam on quantum physics many students gave different realization of the mechanical part of the setup. The best solution with longest steps of the conductivity was given by spring shaped wires with very low frequency of vibration and small velocity of connection and disconnection. 
Our concrete conclusions is that just this setup gives the simplest method for measurement of Planck constant by electronic phenomenon. 
Definitely the simplicity of the setup will have relevance for physics teaching 
of quantum phenomena and learning at the university and even in high school level.

\section{\label{sec:ack}Spelling of authors contribution and acknowledgments}
This work is partially fulfilled requirement for the exam on Thermodynamics and Statistical Physics, lecturer T.~Mishonov, of the students D.~Damyanov, I.~Pavlova, S.~Ilieva. The first prototype of the setup is made by V.~Yordanov and T.~Mishonov. The students, supervised by V.~Gurev, have realized their setup from scratch and now it is in the Atomic physics teaching laboratory. The text in this composcript was completely written by the students. They would like to express their gratitude to T.~Velchev, N.~Pavlov, S.~Damyanov and I.~Iliev for the support. The authors appreciate many creative suggestions made by colleagues with critical reading of the manuscript. 


\newpage

\begin{center}

\large{\textbf{Translation in Bulgarian:\\Измерване на константата на Планк чрез квантуване на Ландауер\\ за студентски лаборатории}}

\medskip \medskip

\normalsize

Десислав С Дамянов, Илияна Н Павлова, Симона И Илиева, \\
Васил Н Гурев, Васил Г Йорданов, Тодор М Мишонов

\emph{Физически факултет, Софийски университет Св. Климент Охридски,\\
бул.~Джеймс Баучер 5, 1164, София, България}

\medskip \medskip

\end{center}

\leftskip=3cm \rightskip=3cm \noindent
\textbf{Резюме:}
Опростена експериментална постановка за измерване на константата на Планк използвайки квантуване на Ландауер на проводимостта между две докосващи се златни жички е описана.
Тя се състои от две златни жички с дебелина 500~$\mu$м и 1.5~см дължина, и операционен усилвател.
Постановката струва по-малко от \$30 и може да бъде реализирана във всяка учебна лаборатория за две седмици.
Изисква се употребата на осцилоскоп.

\leftskip=0pt\rightskip=0pt

\medskip

\section{Въведение}

Целта на настоящата работа е да опише експериментална постановка за наблюдение на квантуване на Ландауер в проводимостта и определянето на константата на Планк~\cite{Planck}, която може лесно да се реализира във всяка учебна лаборатория.

Квантуването на проводимостта е предсказано от Ролф Ландауер през 1957~\cite{IBM}.
Вижте например изследванията на Ландауер~\cite{Landauer1989} и Бютикер~\cite{Buttiker}.

През последните няколко години наблюдението на квантовата проводимост и нейното използване за определянето на константата на Планк се превърна в стандартна експериментална работа за студенти~\cite{Tolley,Foley,conference,Soukiassian}.

За повече информация относно историята за измерването на константата на Планк, виж например изследването на Щейнер~\cite{Steiner}.

Квантуването на проводимостта има елементарно обяснение, подходящо за гимназиални ученици, което може да бъде считано за илюстрация на модела на Бор за водородния атом \cite{Stoev}.

Експерименталната постановка е достъпна в учебната лаборатория на атомна физика.
Както е описано в други работи~\cite{Tolley,Foley}, намирането на начин за свързване и разкачване на жичките е оставено на въображението на студентите.
Въпреки това, експерименталната постановка е здрава и повтаряема във времевата рамка на три часа (продължителността на лабораторен клас).
Ако се използват съединения с пукнатини (break junctions)~\cite{Soukiassian}, измерването става по-надеждно, но вдъхновението от експерименталната тънкост се губи.

Най-повтаряемият експеримент по квантова проводимост се извършва в технологични структури без атомно движение.
Единственият регулатор на квантовите канали е усилването по напрежение~\cite{Wees,Kouwevhoven,Kouwenhoven1992}.

\section{Теоретичен модел}

\subsection{Проводимост на едномерен метал, изключвайки разсейване}

\begin{figure}[h]
\includegraphics[width=0.47\textwidth]{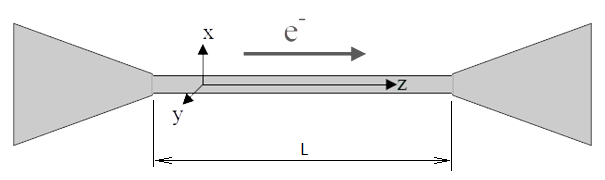}
\caption{Картинка на Ландауер за едномерен транспорт~\cite{IBM,Landauer1989}.
Едномерен (1D) електронен вълновод между два масивни метални електрода.
Различни трансверсални моди отговарят на различни квантови канали на проводимост.\cite{Kouwenhoven1992}
Средният свободен пробег е по-голям от дължината на вълновода $L$ и електроните балистично прелитат през контактната площ между електродите.
В експерименталната реализация на настоящата работа електродите са две златни жички, докато вълноводът е просто точката на контакт между тях cf.~\cite{Tolley}.}
\label{fig1}
\end{figure}

Според разсъждението на Ландауер~\cite{IBM,Landauer1989},проводник може да бъде разглеждан като едномерна система от свободни електрони, схематично представени на \Fref{fig1}, ако дължината $L$ е много по-малко от средния свободен пробег.

Разпределението на електроните по свободните енергийни състояния в система с голям брой идентични частици се описва със статистика на Ферми-Дирак~\cite{Fermi}.
Разпределението на Ферми-Дирак дава средния брой фермиони с импулс $p$ в едно-частичково състояние~\cite{Reif}
\begin{equation}\label{}
n_p=\frac{1}{exp[(\epsilon _p - \mu)/k_BT^{\prime}]+1},
\end{equation}
където $k_B$ е константата на Болцман, $T^\prime$ е абсолютната температура, $\epsilon_p$ е енергията на едно-частичковото състояние, което е известно, че е равно на $\epsilon_p=p^2/2m$, $\mu$ е химичният потенциал и $p$ е импулсът на електрона.
При нулева температура химичният потенциал е сума от енергията на Ферми и потенциалната енергия на един електрон~\cite{Blakemore,Kittel}.

Има две възможно стойности за $n_p$
\begin{equation}
n_p=
\begin{Bmatrix}
1 \textrm{ in case of } \epsilon _p < \mu\\
0 \textrm{ in case of } \epsilon _p > \mu 
\end{Bmatrix},
\end{equation}
които се извеждат от принципа на Паули.

Сега представяме тока като поток от електрони, използвайки сумиране
\begin{equation}\label{}
I=q_e\sum_{\alpha,p;\;(v_p>0)}v_p\frac{\bar{n_p}}{L},\qquad v_p=\partial _p\epsilon _p=\frac{p}{m},
\end{equation}
където $q_e,\alpha$ и $v$ са съответно: зарядът, спинът и скоростта на електрона.
Предполагаме, че $q_e$ е известен.
Величината $\bar{n_p}/L$ е осреднена пространствена плътност на електрони, имащи импулс $p$ и $q_e \bar{n_p}/L$ е електричната плътност, и показва средния брой частици на единица дължина.
Неявно в разсъждението на Ландауер се предполага балистично разпространение на свободни частици в периодични гранични условия с геометричен период $L$.

Можем да преобразуваме сумирането в интегриране, използвайки приближението\cite{Laurendeau}
\begin{equation}\label{}
\sum_{p}=\frac{L^D}{(2\pi\hbar)^D}\int_{0}^{p_F}d^Dp,\qquad v=\frac{p}{m}\geq 0,
\end{equation}
където $D$ е пространственото измерение, $D=1$ в нашия случай, $\hbar$ е редуцираната константа на Планк, $L^D$ е разглежданият обем (в този случай е дължината на проводника) и $p_F$ е импулсът на Ферми.
Формално сумиране може да бъде заместено с интегриране само в границата $L\rightarrow\infty$.

Следователно, токът може да бъде записан като
\begin{equation}\label{eq:current}
I=2q_eL\int_{0}^{p_F}\frac{p}{m}\frac{dp}{2\pi \hbar}\frac{1}{L},
\end{equation}
където първият множител 2 взема предвид сумирането по спин.
Скоростта е представена като $p/m$ и $h=2\pi\hbar$.

След интегриране, уравнението се превръща в
\begin{equation}\label{}
I=\frac{2q_e}{2\pi \hbar}\left (\frac{p^{2}}{2m}\right)\bigg\vert_{0}^{p_F}=\frac{2q_e}{2\pi \hbar}E_F=\frac{2q_e}{2\pi \hbar}q_eU=\frac{2q_e^{2}U}{h}=\sigma_0 U,
\end{equation}
където $\sigma_0$ е кванта проводимост~\cite{Soukiassian}
\begin{equation}
\label{eq:sigma}
\sigma_0 = \frac{q_e^2}{\pi \hbar}
=\frac{2}{R_\mathrm{H}}
=77.5\,\mathrm{\mu S}=\frac{1}{12.9 \,\mathrm{k\Omega}},
\end{equation}
където $R_\mathrm{H}=2\pi\hbar/q_e^2=25812.8074434(84)\,\Omega$~\cite{Mohr}
е квантуваното Холово съпротивление (константа на фон Клицинг), свързана с метрологичната дефиниция на единицата за съпротивление $\Omega.$
Струва си да се отбележи, че в уравнение \Eqref{eq:sigma} ефективна маса на квазичастици $m$ се съкращава и този резултат е приложим за всички 1D проводници.
Дължината $L$ е също без значение и квантуване на проводимостта може да бъде видяна не само в 1D електронни вълноводи, но дори и в точкови контакти, за които $L=0$, виж \cite{Landauer1989,Kouwevhoven}.
За приложимост на квантуване на Ландауер за докосващи се жички, необходимо за размера на контакта (от порядъка на няколко \r{A}) е да бъде много по-малък от средния свободен пробег в метали, който е от порядъка на 100~\r{A}.
Електроните трябва да прелетят през контактната площ като свободни частици.
Ето затова, тънък окисен слой мед може да размие квантуването на проводимостта и е по-добре да се използват златни жички.

Квантуването на проводимостта е също свързана със скоростта на Бор
\begin{eqnarray}
&&\frac{1}{4\pi \epsilon _0R_\mathrm{H}}
=\frac12\,\frac{\sigma _0}{4\pi \epsilon _0} = \frac{v_\mathrm{_{Bohr}}}{2\pi}, \\
&&v_\mathrm{_{Bohr}} = \frac{e^2}{\hbar }
=\alpha_\mathrm{_{Zommerfeld}}c,
\quad e^2 = \frac{q_{e}^2}{4\pi\epsilon_0}.\nonumber
\end{eqnarray}
Всички формули са написани в система SI.
В система CGS  $4\pi\epsilon_0=1$, и проводимостта има своята естествена размерна скорост~\cite{Parsell}.
В система Heaviside-Lorentz $\epsilon_0=1=c$.

Има по-лесен метод да се изчисли интегралът за тока в \Eqref{eq:current}, използвайки само сумиране на аритметична прогресия~\cite{Stoev}, която е подходяща за гимназиални ученици.

\subsection{Формализъм на Ландауер}
\label{s22}

По-реалистичен случай в наблюдението на тока преминаващ през проводник е когато разсейването се отчита.

Разглеждаме проводникът като двумерна (2D) система.
Електроните са затворени от безкраен 2D потенциал в $x$- и $y$-направленията.
В този случай, трябва да използваме квантова механика, в частност уравнението на Шрьодингер за частица в кутия.
Неговото решение дава дискретен брой собствени състояния, които още се наричат моди (или канали, ако се разглежда проводимост).
Пълната енергия на проводника може да бъде получена като сума на енергията на страничната мода  и енергията на едномерното решение в направление $z$.

Използвайки Ландауеровия формализъм за ток течащ през малки стеснения, може да бъде изведено следното
\begin{equation}
\label{}
\sigma =\sigma _0\sum_{i=1}^{N}T_i,
\end{equation}
където $\sigma _0$ е квантовата проводимост, $N$ е броят на проводящите канали и $T_i$ е коефициентът на предаване на $i$-тия канал в жичката~\cite{Landauer1989,Soukiassian}.
Множителите $T_i$ са важни, тъй като проводникът не е идеален.
Те показват вероятността електрон да премине през стеснението, преминавайки през $i$-тия канал.
$T_i$ се различава от 1, когато разсейването назад стане значимо в транспортните процеси.
Ако дължината на съединенията с пукнатини (break junctions) стане по-малка от средния свободен пробег на електрон в метал ($\sim$~100~\r{A}), може да се приеме че $T_i=1$ за всички канали.
Формулата на Ландауер става
\begin{equation}\label{}
\sigma =\sigma _0N\textrm{  or } \frac{\sigma }{\sigma _0}=N,
\end{equation}
тоест проводимостта разделена на квантовата проводимост винаги е равна на цяло число~\cite{Soukiassian}.

В следващата секция описваме експерименталната реализация.

\section{Експериментална част}

\subsection{Електрична схема}
\label{s31}

\begin{figure}[h]
\includegraphics[width=0.6\textwidth]{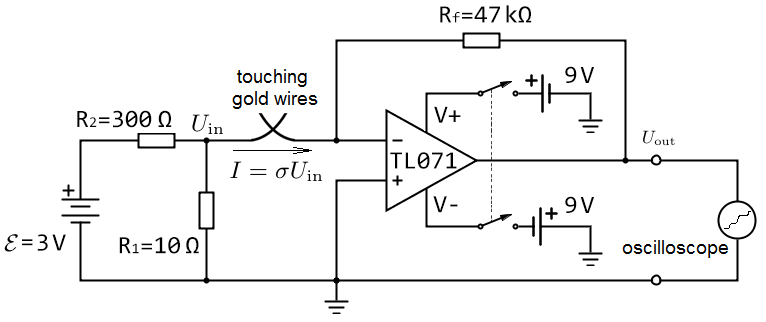}
\caption{Електрична схема за измерване на константата на Планк.
Преобразувателят ток-в-напрежение образуван от операционният усилвател TL071 и резисторът в обратната връзка $R_\mathrm{f}$ има изходно напрежение пропорционално на входния ток $U_\mathrm{out}=-R_\mathrm{f} I$. 
Входният ток, който тече през златните жички се дава от проводимостта $I=\sigma U_\mathrm{in}$, където $U_\mathrm{in}$ е падът на напрежение на $R_1$ от делителя на напрежение образуван от $R_1$ и $R_2$.
Изходното напрежение е пропорционално на проводимостта на златните жички $U_\mathrm{out}=-\sigma R_\mathrm{f} U_\mathrm{in}$ и се записва с цифров осцилоскоп.
Алтернативен анализ на схемата е, че входно напрежение $U_\mathrm{in}=\mathcal{E}R_1/(R_1+R_2)$ се усилва от инвертиращ усилвател с резистора в обратна връзка $R_\mathrm{f}$ и входен резистор $1/\sigma$.
Коефициентът на усилване е $G(t)=-\sigma R_\mathrm f$ и осцилоскопът показва време-зависимото изходно напрежение $U_\mathrm{out}(t)=G U_\mathrm{in} \propto \sigma(t).$
По този начин можем да видим квантуваната време-зависима проводимост между докосващите се златни жички.}
\label{fig2}
\end{figure}

Експериментът се извършва като се използва електрическата схема показана на \Fref{fig2}, където кръстосаните линии представят две златни жички.
Тази схема е подобна на използваната във Foley et al.\cite{Foley}

За наблюдение на времева зависимост на проводимостта и квантуването му, токът минаващ през проводящите жички се измерва, използвайки ток-в-напрежение преобразувател, както е показано на \Fref{fig2}.

Когато златните жички образуват един квант проводимост, те могат да бъдат представени като резистор $r_1=12.9 \,\mathrm{k\Omega}$.
В този случай, \Fref{fig2} показва схемата на инвертиращ усилвател, който измерва входното напрежение $U_\mathrm{in}$ от делителя на напрежение образуван от $R_1$ и $R_2$.
За стойностите на $R_1=10\, \mathrm{\Omega}$ и $R_2=300\, \mathrm{\Omega}$, падът на напрежение на $R_1$ от делителя е приблизително 100~mV.
Стойността на $R_1$ е избрана така, че делителят на напрежение да не е натоварен от входното съпротивление на инвертиращия усилвател за първите десет от квантовите съпротивления $R_1 \ll r_N, N=1,\dots, 10$.
Стойността на $R_2$ е избрана така, че изходното напрежение от делителя $U_\mathrm{in}$ да е по-малко от максималното изходно напрежение на инвертиращия усилвател, разделено на усилването $G_n$ за първите десет от квантовата проводимост $U_\mathrm{in} = 3\, \mathrm{V} \times R_1/(R_1+R_2) < 9\, \mathrm{V} / G_N , N=1,\dots, 10$. 

Усилването по напрежение на инвертиращ усилвател е дадено като
\begin{equation}\label{}
G_N=\frac{U_\mathrm{out}}{U_\mathrm{in}}=-\frac{R_\mathrm{f}}{r_N}
\end{equation}
Използваният операционен усилвател е TL071.
Той е много евтин, широко разпространен и достъпен в Dual In-Line Package (DIP).
Има ширина на честотната лента 3~MHz съответстваща на усилване $G$=1.
В случая на един квант проводимост, усилването е $G_1 = \frac{R_\mathrm{f}}{r_1}=3.7$.
Тъй като произведението на ширината на лентата на операционен усилвател с неговото усилване е постоянно, ширината на лентата на използвания инвертиращ усилвател е $B=$0.8~MHz.
Това е много по-малко от ширината на лентата на използваните осцилоскопи  \cite{DS5042M,DS1052E}, което ги прави подходящи за измерване на изходното напрежение $U_\mathrm{out}$.
Ширината на лентата на инвертиращия усилвател е също достатъчно висока, така че можем да измерим първите няколко квантови стъпала с дължина по-голяма от $1/B$=1.25~$\mu$s.
Това е причината да се избере сравнително малка стойност за $R_\mathrm{f}$ от порядъка на няколко десетки $\mathrm{k\Omega}$, така че усилването също се задържа малко и ширината на лентата е висока.
Целта на експеримента е да се определи константата на Планк чрез използване на стойностите на квантуваната проводимост между две докосващи се златни жички.
Тази проводимост и съответстващото съпротивление са свързани с константата на Планк
\begin{equation}\label{r_N}
r_N=\frac{1}{\sigma }=\frac{1}{N\sigma _0}=\frac{1}{2q_e^2} \frac{h}{N}.
\end{equation}
Съпротивлението $r_N=-\frac{U_\mathrm{in}}{U_\mathrm{out}}R_\mathrm{f}$ се получава чрез измерването на стойността на $U_\mathrm{out}$, съответстваща на $N$-тото квантово ниво, и вземайки предвид стойностите на входното напрежение $U_\mathrm{in}$ и съпротивлението в обратната връзка $R_\mathrm{f}$.
Следователно константата на Планк е
\begin{equation}\label{eq:hconst}
h = 2\,q_e^2\,r_N N = -2\,q_e^2\,R_fN \,\frac{U_\mathrm{in}}{U_\mathrm{out}}.
\end{equation}
Тя е изразена чрез физични константи и експериментално определяеми параметри и променливи на постановката.

\subsection{Изработване на постановката}

\begin{figure}[h]
\includegraphics[width=0.6\textwidth]{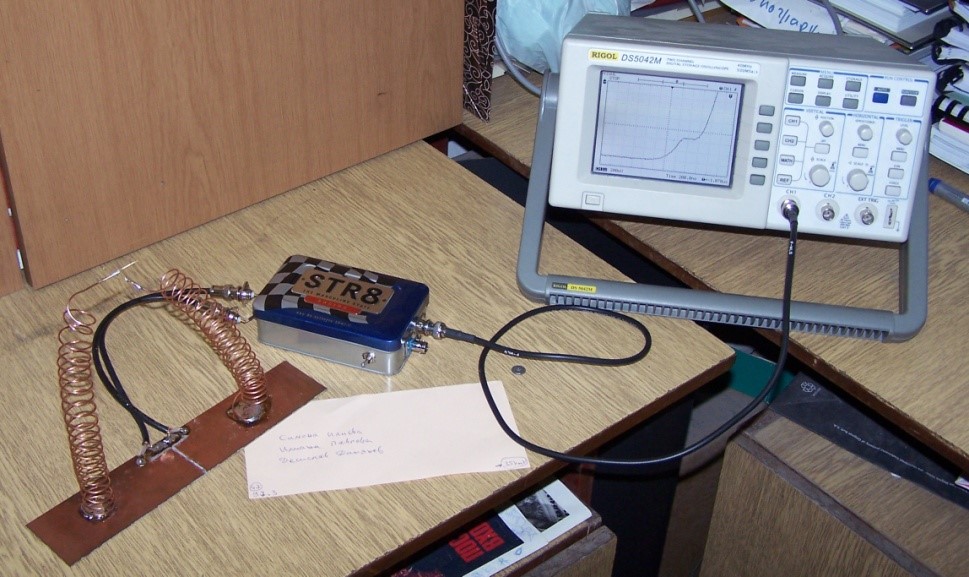}
\caption{Изглед на цялата експериментална постановка.
Вижда се осцилоскопа, показващ типично квантувано стъпало на проводимостта.
Електричната част е в метална кутия, показана с отворен капак в по-голям детайл на \Fref{fig4}.
Механичната част, осигуряваща бавното докосване на металните жички е значително увеличена на \Fref{fig5}.
Всички части на постановката са свързани с коаксиални кабели, терминирани от BNC конектори.}
\label{fig3}
\end{figure}

\begin{figure}[h]
\includegraphics[width=0.47\textwidth]{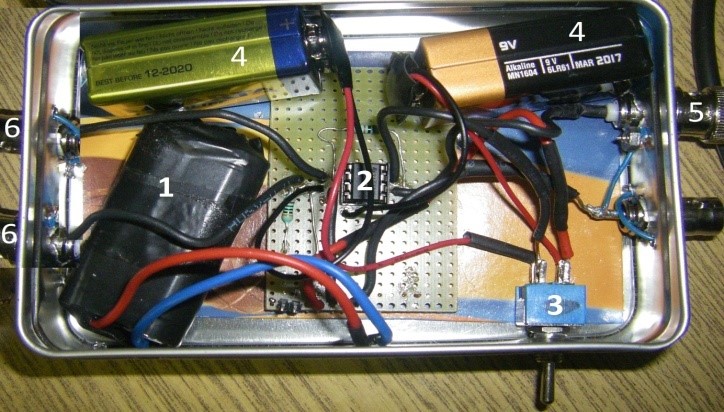}
\caption{Електрична част в металната кутия от \Fref{fig3}.
Могат да се видят:
(1) двете опаковани в черен изолирбанд батерии AA от 1.5~V, всяка даваща електродвижеща сила  $\mathcal{E}$,
(2) операционен усилвател TL071~\cite{TL071}, 
(3) двоен ключ за захранването на операционния усилвател,
(4) две 9~V батерии на захранването на операционния усилвател,
(5) BNC конектор към осцилоскопа,
(6) BNC конектори за коаксиалните кабели, свързани към златните жички, дадени в значително увеличение на \Fref{fig5}.
Може също да се види: изолаторната пластина на дъното на металната кутия, стандартна проектна платка с дупки, отляво на платката единия резистор от делителя на напрежение и резисторът на обратната връзка $R_\mathrm{f}$ над операционния усилвател, и някои несъществени детайли.}
\label{fig4}
\end{figure}

\begin{figure}[h]
\includegraphics[width=0.47\textwidth]{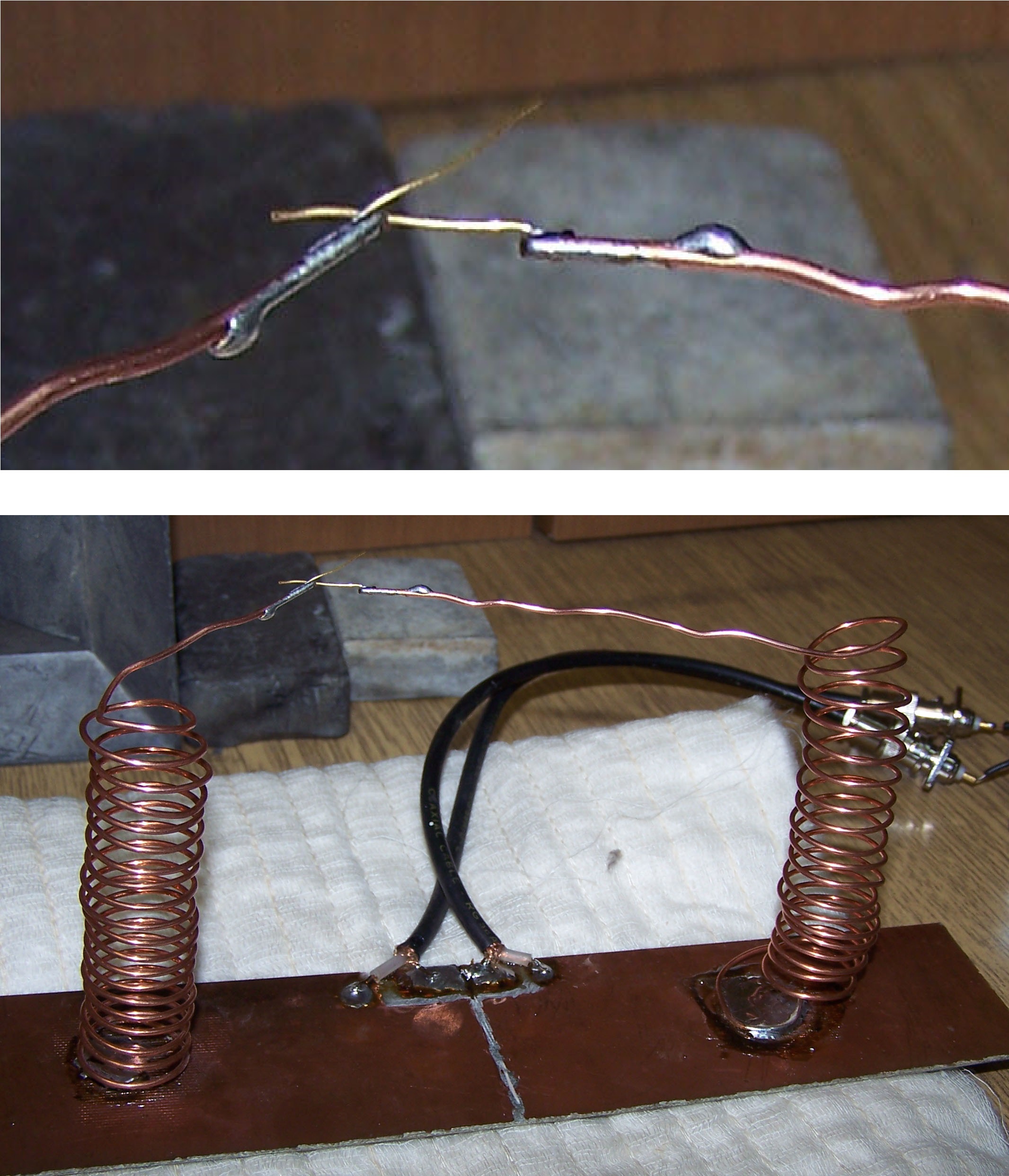}
\caption{Механичната част на постановката.
Къси златни жички (15~мм дълги и 0.5~мм дебели) са запоени на върха на 1.5~мм дебели медни пружини.
Златните жички (увеличени в горната част на фигурата) са в механичен контакт, както схематично е показано на \Fref{fig2}.
Долният край на пружините са запоени на отделени медни пластини.
За медните пластини са запоени проводниците на два коаксиални кабела, които намаляват външния електричен шум.
Екранировките на коаксиалните кабели са запоени към общата точка (земя) на схемата на \Fref{fig2}.
Целта на пружинно-оформените медни проводници и да намали честотите на собствените механични моди на трептения.
Това е механичен филтър за високо-честотен механичен шум и едновременно увеличава времето за наблюдение на квантуваните стъпала на проводимостта.
Когато тези стъпала са достатъчно дълги, можем да използваме ниско-честотни операционни усилватели и осцилоскопи.}
\label{fig5}
\end{figure}

Постановката на \Fref{fig3} се състои от механична и електрична част, които са свързани с коаксиални кабели.
Електричната и механична части са показани съответно на \Fref{fig4} и \Fref{fig5}.

Механичната част е направена по такъв начин, че да предоставя бавно движение на златните жички когато образуват или прекъсват електричен контакт. 
Златните жички са запоени на края на големи пружини (виж \Fref{fig5}).
Жичките са направени от търговско 24-каратово злато за бижута.
Те са 15~мм дълги и имат дебелина 0.5~мм.
Когато пружините осцилират, златните жички се докосват и разделят.
Всяка пружина е 1.5~мм меден проводник с дължина 9~см.
Има 20 навивки с диаметър 2~см.
Пружините са запоени на медна плочка с разделени по средата слоеве за образуване на две индивидуални проводящи повърхности.

Електричната схема, описана в секция \ref{s31}, е реализирана в синята метална кутия показана на \Fref{fig4}.
Източникът на право напрежение е в 3~V държач за батерии, състоящ се от две батерии от 1.5~V всяка.
Операционният усилвател TL071 се захранва от две батерии от 9~V.
Ключът прекъсва захранването на операционния усилвател.

Електричната част е свързана с механичната част и осцилоскопа с 30~см дълги коаксиални кабели.
BNC конекторите на кутията отговарят на характеристичния импеданс на кабелите, който е $50 \,\mathrm{\Omega}$. 

Синята метална кутия от STR8 служи за Фарадеев кафез да потиска електромагнитна интерференция от външната среда.

\subsection{Измерване}

В началото, ключът трябва да бъде включен, за да се захрани операционния усилвател.
Тогава настройваме времевата скала на осцилоскопа, скалата за напрежение и нивото на тригериране (за повече информация виж секция \ref{s5}).

След леко побутване, пружинно-оформените проводници започват да вибрират и златните жички влизат в и излизат от контакт.
Когато жичките се свържат, електрони преминават от една от жиците до другата през $N$ отворени квантови канали (виж секция \ref{s22}).
Едновременно на отделянето на жичките, броят моди намалява и същото прави изходното напрежение.
Резултатният сигнал е поредица от стъпки, поради квантуването на проводимостта.
Височината на всяко стъпало отговаря на целочислен квант на проводимостта.
Стойността на $U_\mathrm{out}$, отговаряща на всяко стъпало, се измерва като се използва осцилоскопа.
Входното напрежение на инвертиращия усилвател $U_\mathrm{in}$ се измерва директно с волтметър като пад на напрежение на резистора $R_1$.
Съпротивлението $R_\mathrm{f}$ се измерва с омметър.
Тогава, константата на Планк се пресмята използвайки \Eqref{eq:hconst}  и измерените стойности за  $U_\mathrm{out}$, за избрано $N$-то стъпало, $U_\mathrm{in}$ и $R_\mathrm{f}$.

При извършване на експеримента са възможни два подхода.
Първият е да се разделят двете жички.
Вторият е да се докоснат в контакт.

При първия подход, ако жичките са свързани в началото на измерването, изходното напрежение е около -9~V, тъй като $r_N=0$ и усилването на инвертиращия усилвател е близо до своето усилване при отворена верига (open-loop gain) и $U_\mathrm{in}$ се увеличава до захранващото напрежение.
В този случай тригер нивото на осцилоскопа се настройва около 700~mV под него.
Квантуването на проводимостта се наблюдава по време на разделянето на жичките.

При втория подход, ако жичките са първоначално отделени, изходното напрежение е около 0~V, тъй като усилването на инвертиращия усилвател е нула и също е откачен от входния източник.
В този случай тригер нивото се настройва около 700~mV над нулевото ниво, което отговаря на изходно напрежение $U_\mathrm{out}$, когато има няколко кванти на проводимост.
Квантуването на проводимостта се наблюдава, когато жичките се докоснат.

И двата метода са подходящи за измервания и изглежда да са еднакво ефективни.

\section{Резултати}

Входното напрежение се измерва $U_\mathrm{in}=97.3\,\mathrm{mV}$ и не се променя по време на експеримента.
Измерената стойност на $R_\mathrm{f}=(47.0\pm0.1)\,\mathrm{k}\Omega$.
Точността на тази стойност е много по-висока от точността на измерванията с осцилоскопа, затова грешката на $R_\mathrm{f}$ е без значение.

\begin{figure}
\includegraphics[width=0.6\textwidth]{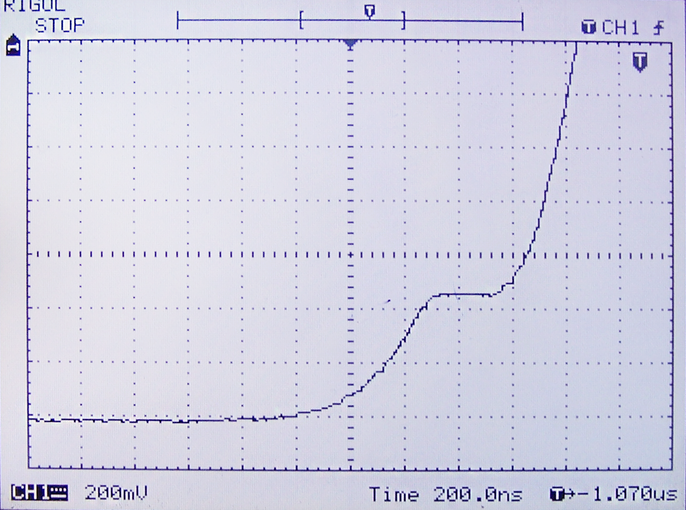}
\caption{Напрежение пропорционално на проводимостта като функция на времето.
Лесно се вижда стъпало, на което проводимостта е равна на кванта проводимост $\sigma_0.$ 
Режим на разделяне на златните жички.
Осцилоскоп с 40~MHz ширина на честотната лента е използван.
Оста X е настроена на 200~ns и оста Y е на 200~mV.
Стъпалото с височина 440~mV и дължина от 200~ns отговаря на една единица квантова проводимост.
Ширината на стъпалото с квантувана проводимост е около 200~ns.}
\label{fig6}
\end{figure}

\begin{figure}
\includegraphics[width=0.6\textwidth]{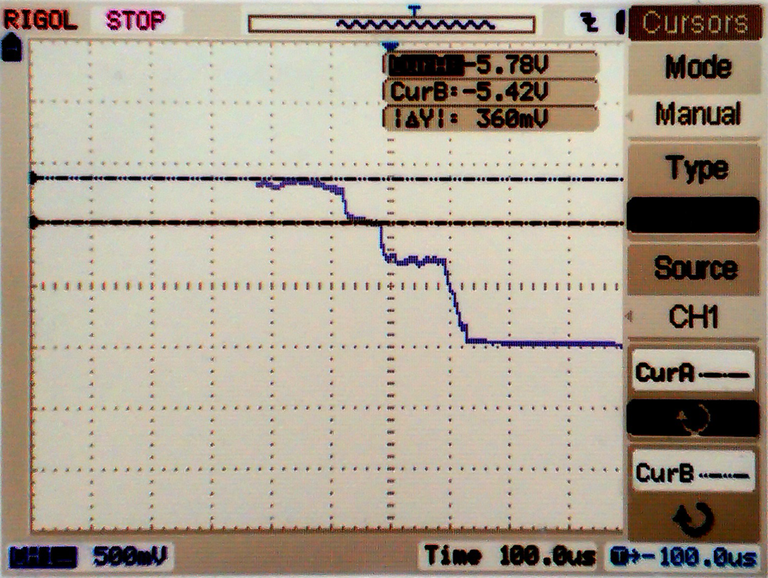}
\caption{Три стъпала на квантова проводимост като функция на време могат да се видят в режим на свързване на златните жички.
Осцилоскоп с $50 \,\mathrm{MHz}$ ширина на лентата е използван.
Оста X е настроена на $100\,\mathrm{\mu s}$ и оста Y e на 500~mV.
Стъпалото между маркерите с височина от 360~mV и дължина 60~$\mu$s отговаря на една единица квантова проводимост.
Хоризонталните курсори описват избраното ниво на проводимост.
Отклонението на кривата от хоризонталното стъпало на квантуваната проводимост с повече информативно от лента за грешка около експериментална точка.
Да отбележим, че ширината на стъпалата 50--100~$\mu$s е значително по-голяма от ширината на стъпалата показани на \Fref{fig7}.}
\label{fig7}
\end{figure}

\Fref{fig6} и \Fref{fig7} показват снимки на екрана на осцилоскопа.
Те илюстрират изходното напрежение на операционния усилвател като функция на времето, използвайки два различни осцилоскопа.
\Fref{fig6} показва експерименталните криви, когато измерването е направено в режим на отделяне на жичките използвайки Rigol~DS5042M цифров осцилоскоп.
Кривите на \Fref{fig7} са показани в режим на свързване използвайки Rigol~DS1052E цифров осцилоскоп.
Броят на отворените квантувани канали е тема на интерпретация.
Апроксимираме до най-близкото цяло число $N$.
Резултати от измерванията са представени на в Таблица~\ref{t1}.
\begin{table}[!ht]
\centering
\begin{center}
\begin{tabular}{@{\hspace{10pt}}c@{\hspace{5pt}}c@{\hspace{5pt}}c@{\hspace{5pt}}}
\hline
\hline
  изходно  & брой  & константа  \\
  напрежение & канали & на Планк \\
   $U_\mathrm{out}$ [mV]   & N & {$h\,[10^{-34}$J\,s]}   \\
\hline
 720 & 2 & 6.522 \\
 760 & 2 & 6.179 \\
 1800 & 5 & 6.522 \\
 360 & 1 & 6.522 \\
 440 & 1 & 5.336 \\
 360 & 1 & 6.522 \\
 320& 1 & 7.337 \\
 360& 1 & 6.522 \\
 320& 1 & 7.337 \\
 720& 2 & 6.522 \\
 700& 2 & 6.708 \\
 360 & 1 & 6.522 \\
 360 & 1 & 6.522 \\
 360 & 1 & 6.522 \\
 360 & 1 & 6.522 \\
\hline
\hline
\end{tabular}
\caption{Получени резултати от измерванията от различни експерименти.
Първата колона е нивото на напрежение на проводящото стъпало с N канали.
Виж например \Fref{fig6} и \Fref{fig7}.
Броят канали е написан във втората колона.
В последната колона е дадена изчислената стойност на константата на Планк.}
\label{t1}
\end{center}
\end{table}

Константата на Планк изчислена като средна стойност и стандартно отклонение на нашите измервания представени в Таблица~\ref{t1} е
\begin{equation}
h=(6.54\pm  0.45)\times 10^{-34} \, \mathrm{J\, s}\,.
\end{equation}
Отклонението от препоръчана стойност от CODATA 2010 ($h = 6.626\,07123(133)\times 10^{-34}\,\mathrm{J\,s}$) \cite{Mohr} е 1.3\%.


\section{Упътвания за студенти и асистенти в учебни лаборатории}
\label{s5}

За да се получи константата на Планк използвайки метода и експерименталната постановка представени в тази работа, студенти трябва да следват инструкциите по-долу.

Включете ключа на металната кутия. 
Ключът прекъсва захранващото напрежение на операционния усилвател.
Сега измерете напрежението веднага след златните жички, използвайки волтметър.
Измерената стойност е входното напрежение $U_\mathrm{in}$ и е приблизително 100~mV.

След това свържете осцилоскопа към електричната част на постановката използвайки BNC конекторите.
Ние използваме Rigol~DS5042M \cite{DS5042M} and Rigol~DS1052E \cite{DS1052E}, но всеки цифров осцилоскоп със степен на семплиране по-голяма от 10-мега семпли за секунда може да се използва.
Тази честота се определя от времевия интервал на квантуваната проводимост.
За нашата експериментална постановка тези времена са от порядъка на $\mu$s и в случай на добра механика с много ниски собствени честоти, времевите интервали могат да бъдат дори по-големи, до 1~ms.

Настройте времевата скала на устройството.
Вземете предвид, че стабилността на механичната част на постановката влияе на зависимостта на напрежението от времето.
В случай на бавно свързване и разделяне на златните жички, X скалата трябва да бъде настроена на 100~$\mu$s/дел. 
В противен случай, X скалата трябва да е 0.2--50~$\mu$s/дел.

Настройте скалата на напрежението. 
Коефициентът на усилване на инвертиращия усилвател е $G_N=-\frac{1}{r_N}\,R_\mathrm{f}=3.7\,N$ за $N$ проводящи кванти.
Тъй като резултатното напрежение е  $U_\mathrm{out}=-G_N\,U_\mathrm{in}$, скалата на напрежение трябва да бъде настроена на 200-500~mV/дел.

Освен това, бърза проверка на надеждността на постановката е препоръчително.
Когато златните жички са устойчиво свързани, хоризонтална линия на около -9~V се появява на екрана на осцилоскопа.
Резултатното напрежение е приблизително равно на захранващото напрежение на инвертиращия операционен усилвател.
Когато жичките се разделят една от друга, резултатното напрежение е 0~V и хоризонтална линия на 0 трябва да е на екрана.

В началото на всяко измерване жичките могат да са свързани или разделени.
Тъй като резултатния сигнал се предполага да бъде последователност от стъпала, ниво на тригериране трябва да се настрои.
Правилният режим на тригериране е ``Edge''.
``Edge'' тригериране се получава, когато входният тригер премине през указано ниво на напрежение на специфицираната посока на наклона.
Обхватът на тригериране трябва да е ``Single''.
По този начин, когато се получи тригериращо събитие, осцилоскопът записва една форма на сигнала и спира.

Ако жичките са първоначално свързани, тригерирането е на качващ наклон.
В случай, че са разделени, тригерирането е на падащ наклон.
Нивото на тригериране трябва да бъде настроено около
$U_\mathrm{trig}=U_\mathrm{out} (N=2)\approx 700 \,\mathrm{mV}$ 
под или над първоначалната линия на двата подхода съответно.
Преподаващите асистенти могат да намерят повече информация за тригериране в ръководствата за употреба на осцилоскопите~\cite{DS5042M,DS1052E}.

При извършването на експеримента, пружините могат да бъдат оформени по различни начини и разнообразни действия могат да бъдат предприети за подобряване на резултатния сигнал.
Възможно е да се промени размера на едната или на двете пружини, да се ограничи тяхното движение, например да се направи едната неподвижна, да се наклони медната плочка, да се сложи мек плат под плочката, да се направят по-големи пружини с по-ниски вибрационни честоти и още много.
На студентите се дава възможност да измислят и проверят техните собствени идеи и да променят пружините както те предпочитат.
Например, студенти от същия курс направиха алтернативна реализация използвайки дълъг лост със закачен на дългото му рамо микрометричен винт.
Когато завъртят винта, късото рамо премества златна жичка, която се свързва или се разделя от друга златна жичка.

Когато желания маниер на движение се постигне, лек натиск на страната по медната плочка или по повърхността на експерименталната маса е достатъчен, за да произведе добър сигнал на екрана на осцилоскопа.
В повечето случаи, много опити се правят преди целта да бъде достигната.
Също е много вероятно само едно стъпало ясно да се вижда на осцилоскопа.
Това означава, няколко графики трябва да се получат, за да се пресметна константата на Планк акуратно.
Щом се графиките се получат, височината на всяко стъпало трябва да бъде сметнато.
Това е изходното напрежение $U_\mathrm{out}$. 

След като всички измервания са направени, ключът трябва да се изключи, за да се спре захранващото напрежение на операционния усилвател.

Накрая, константата на Планк се пресмята използвайки \Eqref{eq:hconst}.
Стойността на елементарния заряд на CODATA 2010 е $1.602176565(35)\times 10^{-19}$~C~\cite{Mohr}.
Също така, зарядът на електрона може да бъде измерван в учебни лаборатории.
За тази цел, постановка направена от студенти е достъпна в учебната лаборатория по Атомна физика.
Определената константа на Планк трябва да бъде сравнена с препоръчаната стойност на CODATA 2010.

\section{Извод}

Бидейки навсякъде в нашето ежедневие, трудно е да се повярва колко полезен контактът между метални проводници може да бъде за измерването на фундаментална физична константа.
Квантуването на Ландауер се оказва изненадващо добра основа за истински експеримент, водещ до като по чудо добри резултати за толкова елементарна експериментална постановка.
Възможността да се наблюдават единици квантова проводимост като стъпала във времевата зависимост на напрежение, създадено от финия допир между две жички, е забележителна.
Години са били необходими да се измери константата на Планк с висока точност, но за учебна лаборатория с осцилоскоп, \$30, две седмици и малка група мотивирани студенти са всичките необходими ресурси за измерване в рамките на 2\%.
Без всякакво съмнение, тази цена е достъпна за всяка учебна лаборатория с желанието да постигне забележими постижения по простота и ефективност.
Тъй като различни подходи са възможни, доза ентусиазъм и прецизност може да се превърнат в горивото за ценно постижение, правейки го за преподаващия асистент заслужаващо да участва в прецизността.

За да си вземат изпита по квантова физика, много студенти дадоха различни реализации на механичната част на постановката.
Най-доброто решение с най-дълги стъпала на проводимостта бе дадено от пружинно-оформените проводници с много ниска честота на вибрация и малка скорост на свързване и разделяне.
Нашият конкретен извод е, че тази постановка дава най-простият метод за измерване на константата на Планк чрез електрично явление.
Определено, простотата на постановката ще има практическо значение за преподаване и научаване на квантови явления както на университетско, така и на гимназиално ниво.

\section{Принос на авторите и благодарности}

Тази работа частично допълва изискването за изпита по Термодинамика и статистическа физика, лектор Т Мишонов, от студентите Д Дамянов, И Павлова, С Илиева.
Първият прототип на постановката бе направен от В Йорданов и Т Мишонов.
Студентите, ръководени от В Гурев, направиха тяхната постановка от нищото и сега тя се намира в учебната лабораторията по Атомна физика.
Текстът в тази статия бе изцяло написан от студентите.
Те биха искали да изразят тяхната благодарност към Т Велчев, Н Павлов, С Дамянов и И Илиев за подкрепата.
Авторите оценяват множество творчески предложения направени от колеги след критични прочитания на ръкописа, а също така и на Алберт Варонов за превода на български език.

\section*{Забележка след превода}

Библиографията не е преведена на български език, 
тъй като всичките цитирани заглавия са на езика, на който са публикувани.
В резюмето на \cite{Foley} е подчертан педагогичния елемент свързан с внасянето на този експеримент от предния фронт на изследванията на физиката на кондензираната материя в учебната студентска лаборатория.
Тази методична статия е публикувана в списание издавано от Американската асоциация на учителите по физика (American Association of Physics Teachers). 
Именно затова е дадена и простата електронна схема за изследване на проводимост, 
която зависи от времето; 
в научните статии електрониката не се коментира като тривиална. 
На страница 391 в \cite{Foley} се коментира, че има множество методи за закрепване на златните жички и студентите сами успяват да изобретят подходящи начини за механичното разделяне и докосване на жичките.
Диаметърът на използваните златни жички в другите експерименти е 0.05~мм,~\cite{Foley} 0.075~мм,~\cite{Tolley}докато в Costa-Kr\"{a}mer et al.~[1995 Surf. Sci. \textbf{342}, LL1144] 
е между 0.1~мм и 1~мм, какъвто е и диаметърът на използваните от нас жички.
Авторите на тези работи \emph{подчертават}, че наноразмерни не са дължината и диаметъра на жичките, а областите на тяхното допиране.

Ученическата версия на теорията е подробно описана в методичната работа,~\cite{Stoev} предшестваща тази университетска версия.

\end{document}